\documentclass[aps,preprintnumbers,superscriptaddress,showpacs]{revtex4}
\usepackage{epsfig}
\usepackage{psfrag}
\usepackage{amsfonts}
\usepackage{graphicx}
\usepackage{dcolumn}
\usepackage{bm}

\begin{document}

\title{$R^2$ corrections to holographic Schwinger effect}

\author{Zi-qiang Zhang}
\email{zhangzq@cug.edu.cn} \affiliation{School of mathematics and
physics, China University of Geosciences(Wuhan), Wuhan 430074,
China}

\author{Chong Ma}
\email{machong@cug.edu.cn } \affiliation{School of mathematics and
physics, China University of Geosciences(Wuhan), Wuhan 430074,
China}

\author{De-fu Hou}
\email{houdf@mail.ccnu.edu.cn} \affiliation{Key Laboratory of
Quark and Lepton Physics (MOE), Central China Normal University,
Wuhan 430079,China}

\author{Gang Chen}
\email{chengang1@cug.edu.cn} \affiliation{School of mathematics
and physics, China University of Geosciences(Wuhan), Wuhan 430074,
China}

\begin{abstract}

We study $R^2$ corrections to the holographic Schwinger effect in
an AdS black hole background and a confining D3-brane background,
respectively. The potential analysis for these backgrounds is
presented. The critical values for the electric field are
obtained. It is shown that for both backgrounds increasing the
Gauss-Bonnet parameter the Schwinger effect is enhanced. Moreover,
the results provide an estimate of how the Schwinger effect
changes with the shear viscosity to entropy density ratio,
$\eta/s$, at strong coupling.
\end{abstract}

\maketitle

\section{Introduction}
It is well known that in quantum electrodynamics (QED) the virtual
particles can turn into real ones when an external strong electric
field is applied. This non-perturbative phenomenon is known as the
Schwinger effect. The production rate $\Gamma$ for a weak-coupling
and weak-field condition has been studied in \cite{JS} long time
ago. Later, it was generalized to the case of arbitrary-coupling
and weak-field regime in \cite{IK}, that is
\begin{equation}
\Gamma\sim exp\big{(}{\frac{-\pi m^2}{eE}+\frac{e^2}{4}}\big{)},
\end{equation}
where $E$ is the external electric field, $m$ is the electron
mass, $e$ is the elementary electric charge. Actually, the
Schwinger effect is not unique to QED but usual for QFTs coupled
to an U(1) gauge field. AdS/CFT, namely the duality between a
string theory in the AdS space and a conformal field in the
physical space-time, can realize a system that coupled with an
U(1) gauge field
\cite{Maldacena:1997re,Gubser:1998bc,MadalcenaReview}. Therefore,
it is of great interest to study the Schwinger effect in the
context of AdS/CFT. The pair productin rate of the $W$ bosons, at
large $N_c$ and large 't Hooft coupling $\lambda\equiv
g_{YM}^2N_c$, was obtained by Semenoff and Zarembo in their
seminal work \cite{GW}
\begin{equation}
\Gamma\sim
exp\big{[}-\frac{\sqrt{\lambda}}{2}\big{(}\sqrt{\frac{E_c}{E}}-\sqrt{\frac{E}{E_c}}\big{)}^2\big{]},
\end{equation}
with
\begin{equation}
E_c=\frac{2\pi m^2}{\sqrt{\lambda}},
\end{equation}
where $E_c$ is the critical electric field. Interestingly, in this
case $E_c$ completely agrees with the DBI result. After \cite{GW},
there are many attempts to address the Schwinger effect in this
direction. For example, the pair production for the general
backgrounds has been investigated in \cite{YS}. The Schwinger
effect in confining backgrounds is discussed in \cite{YS1}. The
potential analysis for Schwinger effect is addressed in
\cite{YS2}. The pair production in constant electric and magnetic
fields has been analyzed in \cite{SB}. Investigations are also
extended to some AdS/QCD models \cite{KH,JS1}. Other related
discussions can be found, for example, in
\cite{SZ,JA,KH1,DD,WF,MG,XW,SC,ZQ1,AS}. For a review on this
topic, see \cite{DK1}.

In general, string theory contains higher derivatives corrections
due to the presence of stringy effect. Although very little is
known about the forms of higher derivative corrections, given the
vastness of the string landscape one expects that generic
corrections can occur \cite{MRD}. Motivated by this, some
quantities have been investigated in conformal field theories dual
to gravity with higher derivative corrections. Such as $\eta/s$
\cite{MB,MB1,YK}, heavy quark potential \cite{JN}, imaginary part
of potential \cite{JN1}, drag force \cite{KB1}, and jet quenching
parameter \cite{ZQ2}.

As the calculation of the holographic Schwinger effect is very
related to string theory, it is natural to consider various
stringy corrections, such as $R^2$ corrections. In this paper, we
would like to analyze how $R^2$ corrections affect the Schwinger
effect. In addition, it was argued \cite{PK} that
$\eta/s\geq1/4\pi$ can be violated in theories with $R^2$
corrections, therefore, the connection between the shear viscosity
and the Schwinger effect in these $R^2$ theories may be an
interesting fact that comes for free in holography. These are the
main motivations of the present work.

The paper is organized as follows. In the next section, we briefly
review the backgrounds with $R^2$ corrections. In section 3, we
perform the potential analysis for the AdS black hole background
with $R^2$ corrections and evaluate the critical electric field
from the DBI action. In section 4, we study the Schwinger effect
in a confining D3-brane background with $R^2$ corrections as well.
The last part is devoted to conclusion and discussion.

\section{setup}
Let us briefly review the backgrounds with curvature-squared
corrections given in \cite{SM}. Restricting the gravity sector in
the $AdS_5$ space, the leading order higher derivative corrections
can be written as
\begin{equation}
S=\frac{1}{16\pi G_5}\int
d^5x\sqrt{-g}[R+\frac{12}{L^2}+L^2(\alpha_1R^2+\alpha_2R_{\mu\nu}R^{\mu\nu}+\alpha_3R_{\mu\nu\rho\sigma}R^{\mu\nu\rho\sigma})],\label{action}
\end{equation}
where $G_5$ is the five dimensional Newton constant,
$R_{\mu\nu\rho\sigma}$ represents the Riemann tensor, $R_{\mu\nu}$
denotes the Ricci tensor, $R$ stands for the Ricci scalar, $L$
refers to the radius of $AdS_5$ at leading order in $\alpha_i$
where one has assumed that
$\alpha_i\sim\frac{\alpha\prime}{L^2}\ll1$. Other terms with
factors of $R$ or additional derivatives can be suppressed by
higher powers of $\frac{\alpha\prime}{L^2}$ \cite{MB}. However, at
leading order only $\alpha_3$ is unambiguous while $\alpha_1$ and
$\alpha_2$ can be arbitrarily varied by a metric redefinition
\cite{MB,YK}. To avoid this problem, one works with the
Gauss-Bonnet gravity in which $\alpha_i$ can be fixed in terms of
a single parameter, $\lambda_{GB}$. The action of Gauss-Bonnet
gravity in four dimensions is given by
\begin{equation}
S=\frac{1}{16\pi G_5}\int
d^5x\sqrt{-g}[R+\frac{12}{L^2}+\frac{\lambda_{GB}}{2}L^2(R^2-4R_{\mu\nu}R^{\mu\nu}+R_{\mu\nu\rho\sigma}R^{\mu\nu\rho\sigma})],\label{action1}
\end{equation}
where $\lambda_{GB}$ is constrained in the range
\begin{equation}
-\frac{7}{36}<\lambda_{GB}\leq\frac{9}{100},
\end{equation}
where the upper range is determined to avoid causality violation
in the boundary \cite{MB1} and the lower bound comes from
requiring the boundary energy density to be positive-definite
\cite{DM}.

The black brane solution of the Gauss-Bonnet gravity is \cite{RG}
\begin{equation}
ds^2=-m^2\frac{r^2}{L^2}f(r)dt^2+{\frac{r^2}{L^2}}d\vec{x}^2+\frac{L^2}{r^2}\frac{dr^2}{f(r)}
\label{metric},
\end{equation}
with
\begin{equation}
f(r)=\frac{1}{2\lambda_{GB}}\big{[}1-\sqrt{1-4\lambda_{GB}(1-\frac{r_h^4}{r^4})}\big{]}\label{f1},
\end{equation}
and
\begin{equation}
m^2=\frac{1}{2}(1+\sqrt{1-4\lambda_{GB}}),
\end{equation}
where $r$ is the radial coordinate describing the 5th dimension.
$r=r_h$ is the event horizon and $r=\infty$ is the boundary. The
plasma temperature is
\begin{equation}
T=\frac{mr_h}{\pi L^2}. \label{T}
\end{equation}

Moreover, $\lambda_{GB}$ can be related to $\eta/s$ by
\cite{MB,MB1,YK}
\begin{equation}
\frac{\eta}{s}=\frac{1}{4\pi}(1-4\lambda_{GB}),\label{eta}
\end{equation}
one can see that $\eta/s>\frac{1}{4\pi}$ is violated for
$\lambda_{GB}>0$. Also, increasing $\lambda_{GB}$ leads to
decreasing $\eta/s$.

\section{AdS black hole background}

We now follow the calculations of \cite{YS2} to study the
Schwinger effect for the background metric of (\ref{metric}). The
Nambu-Goto action is
\begin{equation}
S=T_F\int d\tau d\sigma\mathcal L=T_F\int d\tau d\sigma\sqrt{g},
\qquad T_F=\frac{1}{2\pi\alpha^\prime},
\end{equation}
where $T_F$ is the fundamental string tension. $g$ denotes the
determinant of the induced metric on the string world sheet with
\begin{equation}
g_{\alpha\beta}=g_{\mu\nu}\frac{\partial
X^\mu}{\partial\sigma^\alpha} \frac{\partial
X^\nu}{\partial\sigma^\beta},
\end{equation}
where $g_{\mu\nu}$ is the metric, $X^\mu$ represents the target
space coordinates.

Using the static gauge
\begin{equation}
x^0=\tau, \qquad x^1=\sigma,
\end{equation}
and assuming that the coordinate $r$ depends only on $\sigma$
\begin{equation}
r=r(\sigma),
\end{equation}
one obtains the induced metric $g_{\alpha\beta}$ as
\begin{equation} g_{00}=\frac{m^2r^2f(r)}{L^2}, \qquad
g_{01}= g_{10}=0,\qquad
g_{11}=\frac{r^2}{L^2}+\frac{L^2\dot{r}^2}{r^2f(r)},
\end{equation}
with $\dot{r}=\frac{\partial r}{\partial\sigma}$. Then the
lagrangian density is found to be
\begin{equation}
\mathcal L=\sqrt{\frac{m^2f(r)r^4}{L^4}+m^2\dot{r}^2}.\label{L}
\end{equation}

Now that $\mathcal L$ does not depend on $\sigma$ explicitly, so
the corresponding Hamiltonian is a constant, that is
\begin{equation}
\mathcal L-\frac{\partial\mathcal
L}{\partial\dot{r}}\dot{r}=constant.
\end{equation}

Considering the boundary condition at $\sigma=0$,
\begin{equation}
\dot{r}=0,\qquad  r=r_c,\qquad (r_h<r_c<r_0)\label{con},
\end{equation}
where we have assumed that the probe D3-brane is put at an
intermediate position ($r=r_0$) between the horizon and the
boundary. It was shown that this manipulation can yield a finite
mass \cite{GW}.

To proceed, one finds
\begin{equation}
\frac{f(r)r^4}{\sqrt{f(r)r^4+L^4\dot{r}^2}}=\sqrt{f(r_c)r_c^4},
\end{equation}
with
\begin{equation}
f(r_c)=\frac{1}{2\lambda_{GB}}\big{[}1-\sqrt{1-4\lambda_{GB}(1-\frac{r_h^4}{r_c^4})}\big{]}.
\end{equation}

Then a differential equation is derived
\begin{equation}
\dot{r}=\frac{dr}{d\sigma}=\sqrt{\frac{r^4f(r)[r^4f(r)-r_c^4f(r_c)]}{L^4r_c^4f(r_c)}}\label{dotr}.
\end{equation}

By integrating (\ref{dotr}) the separate length of the test
particle pair is obtained
\begin{equation}
x=\frac{2L^2}{ar_0}\int_1^{1/a}dy
\sqrt{\frac{f_1(y_c)}{y^4f_1(y)[y^4f_1(y)-f_1(y_c)]}}\label{xx},
\end{equation}
with
\begin{equation}
f_1(y)=\frac{1}{2\lambda_{GB}}\big{[}1-\sqrt{1-4\lambda_{GB}(1-\frac{b^4}{a^4y^4})}\big{]},\qquad
f_1(y_c)=\frac{1}{2\lambda_{GB}}\big{[}1-\sqrt{1-4\lambda_{GB}(1-\frac{b^4}{a^4})}\big{]},\label{f1y}
\end{equation}
where we have introduced the following dimensionless quantities
\begin{equation}
y\equiv\frac{r}{r_c},\qquad a\equiv\frac{r_c}{r_0},\qquad
b\equiv\frac{r_h}{r_0}.
\end{equation}

On the other hand, inserting (\ref{dotr}) into (\ref{L}) one gets
the sum of potential energy (PE) and static energy (SE)
\begin{equation}
V_{PE+SE}=2T_Fmr_0a\int_1^{1/a}dy\sqrt{\frac{y^4f_1(y)}{y^4f_1(y)-f_1(y_c)}}.\label{en}
\end{equation}

Next, we calculate the critical electric field. The DBI action is
given by
\begin{equation}
S_{DBI}=-T_{D3}\int
d^4x\sqrt{-det(G_{\mu\nu}+\mathcal{F}_{\mu\nu})}\label{dbi},
\qquad T_{D3}=\frac{1}{g_s(2\pi)^3\alpha^{\prime^2}},
\end{equation}
where $T_{D3}$ is the D3-brane tension.

By virtue of (\ref{metric}), the induced metric $G_{\mu\nu}$ reads
\begin{equation}
G_{00}=-\frac{m^2r^2}{L^2}f(r), \qquad
G_{11}=\frac{r^2}{L^2},\qquad G_{22}=\frac{r^2}{L^2},\qquad
G_{33}=\frac{r^2}{L^2}.
\end{equation}

According to \cite{BZ}, $\mathcal{F}_{\mu\nu}=2\pi\alpha^\prime
F_{\mu\nu}$, one finds
\begin{equation}
G_{\mu\nu}+\mathcal{F}_{\mu\nu}=\left(
\begin{array}{cccc}
 -\frac{m^2r^2}{L^2}f(r) & 2\pi\alpha^\prime E & 0 & 0\\
 -2\pi\alpha^\prime E & \frac{r^2}{L^2} & 0 & 0 \\
0 & 0 & \frac{r^2}{L^2} & 0\\
0 & 0 & 0 & \frac{r^2}{L^2}
\end{array}
\right),
\end{equation}
which yields
\begin{equation}
det(G_{\mu\nu}+\mathcal{F}_{\mu\nu})=-(\frac{r^4}{L^4})^2[m^2f(r)-\frac{(2\pi
\alpha')^2E^2L^4}{r^4}],\label{det1}
\end{equation}
where we have assumed that the electric field $E$ is turned on
along the $x^1$-direction \cite{YS1}.

Inserting (\ref{det1}) into (\ref{dbi}) and setting the D3-brane
at $r=r_0$, one gets
\begin{equation}
S_{DBI}=-T_{D3}\frac{r_0^4}{L^4}\int d^4x
\sqrt{m^2f(r_0)-\frac{(2\pi\alpha')^2E^2L^4}{r_0^4}}\label{dbi1},
\end{equation}
with
\begin{equation}
f(r_0)=\frac{1}{2\lambda_{GB}}\big{[}1-\sqrt{1-4\lambda_{GB}(1-b^4)}\big{]}.
\end{equation}

It is required that the square root in (\ref{dbi1}) is
non-negative
\begin{equation}
m^2f(r_0)-\frac{(2\pi \alpha')^2E^2L^4}{r_0^4}\geq0.
\end{equation}
which leads to
\begin{equation}
E\leq T_F\frac{r_0^2}{L^2}m\sqrt{f(r_0)}.
\end{equation}

Finally, we end up with the critical field $E_c$ in the AdS black
hole background with $R^2$ corrections
\begin{equation}
E_c=T_F\frac{r_0^2}{L^2}m\sqrt{\frac{1}{2\lambda_{GB}}\big{[}1-\sqrt{1-4\lambda_{GB}(1-b^4)}\big{]}}\label{ec}.
\end{equation}
one can see that $E_c$ depends on the temperature as well as the
Gauss-Bonnet parameter.

To move on, we study the total potential. As a matter of
convenience, we define a parameter
\begin{equation}
\alpha\equiv\frac{E}{E_c}. \label{afa1}
\end{equation}

Then from (\ref{xx}) and (\ref{en}) one finds the total potential
$V_{tot}$ as
\begin{eqnarray}
V_{tot}&=&V_{PE+SE}-Ex\nonumber\\&=&2T_Fmr_0a\int_1^{1/a}dy\sqrt{\frac{y^4f_1(y)}{y^4f_1(y)-f_1(y_c)}}\nonumber\\&-&\frac{2T_Fr_0m\alpha}{a}\sqrt{\frac{1}{2\lambda_{GB}}\big{[}1-\sqrt{1-4\lambda_{GB}(1-b^4)}\big{]}}
\int_1^{1/a}dy
\sqrt{\frac{f_1(y_c)}{y^4f_1(y)[y^4f_1(y)-f_1(y_c)]}}\label{V}.
\end{eqnarray}

\begin{figure}
\centering
\includegraphics[width=8cm]{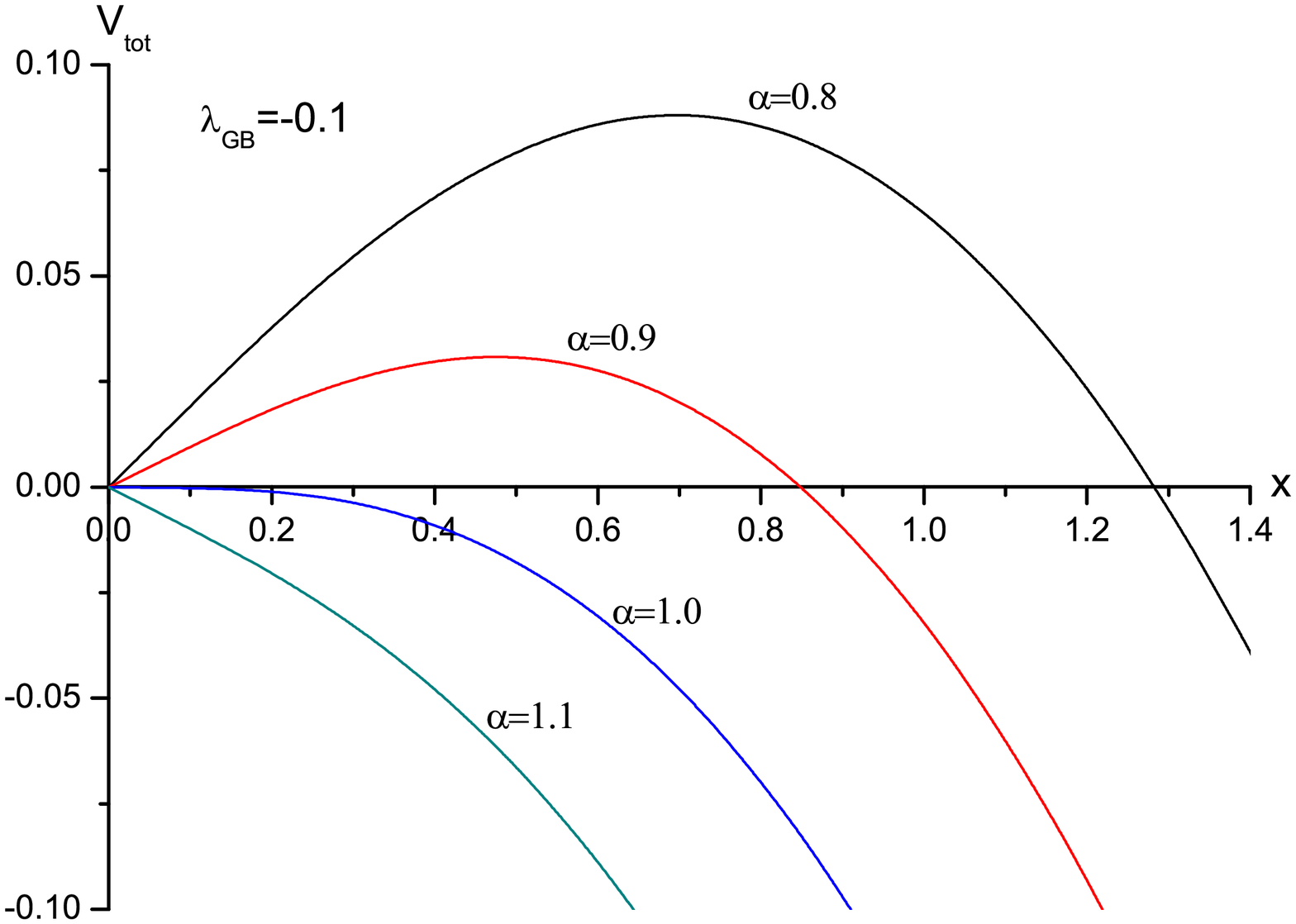}
\includegraphics[width=8cm]{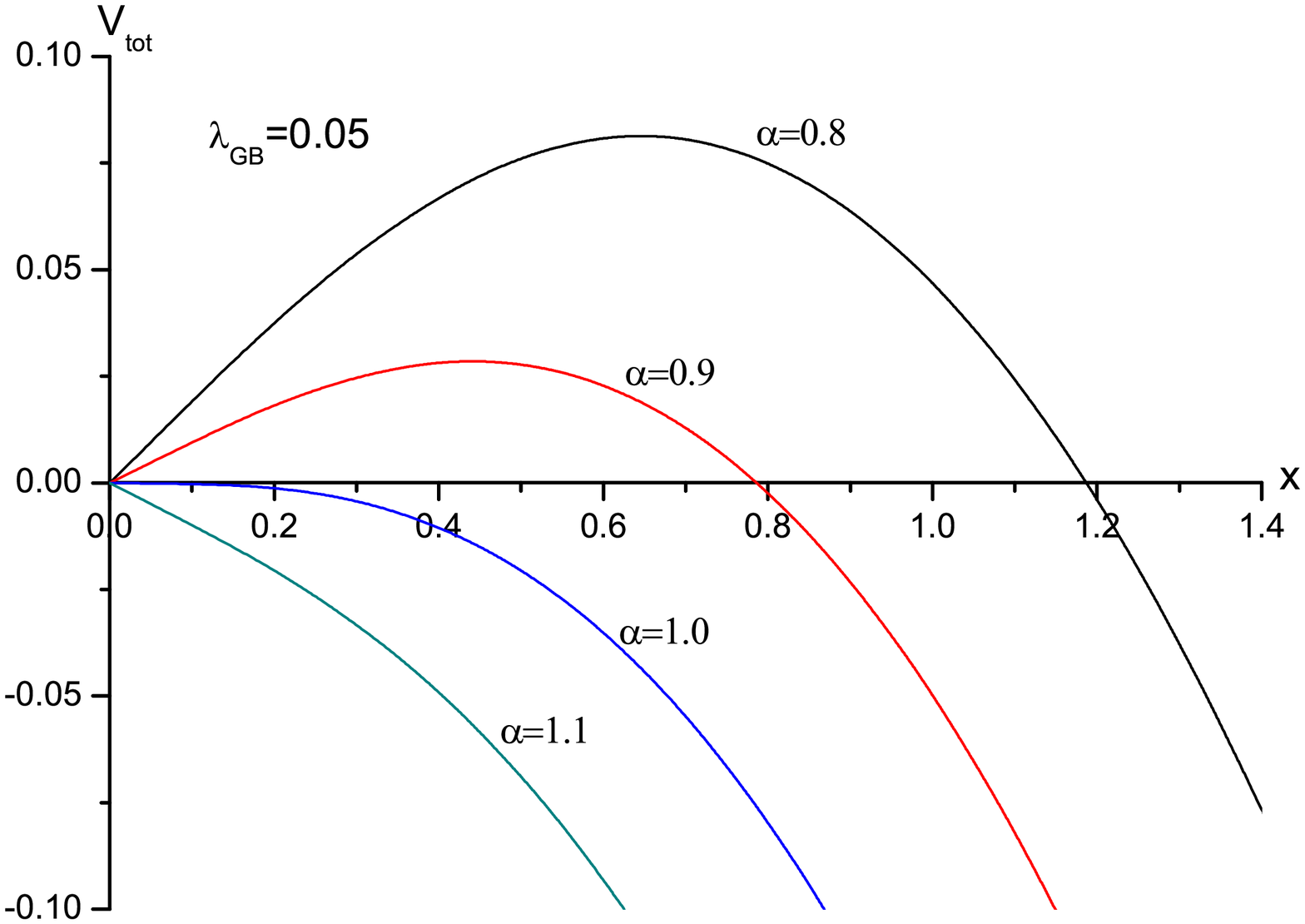}
\caption{$V_{tot}$ against x. Left: $\lambda_{GB}=-0.1$; Right:
$\lambda_{GB}=0.05$. In all of the plots from top to bottom
$\alpha=0.8,0.9,1.0,1.1$, respectively. }
\end{figure}

\begin{figure}
\centering
\includegraphics[width=8cm]{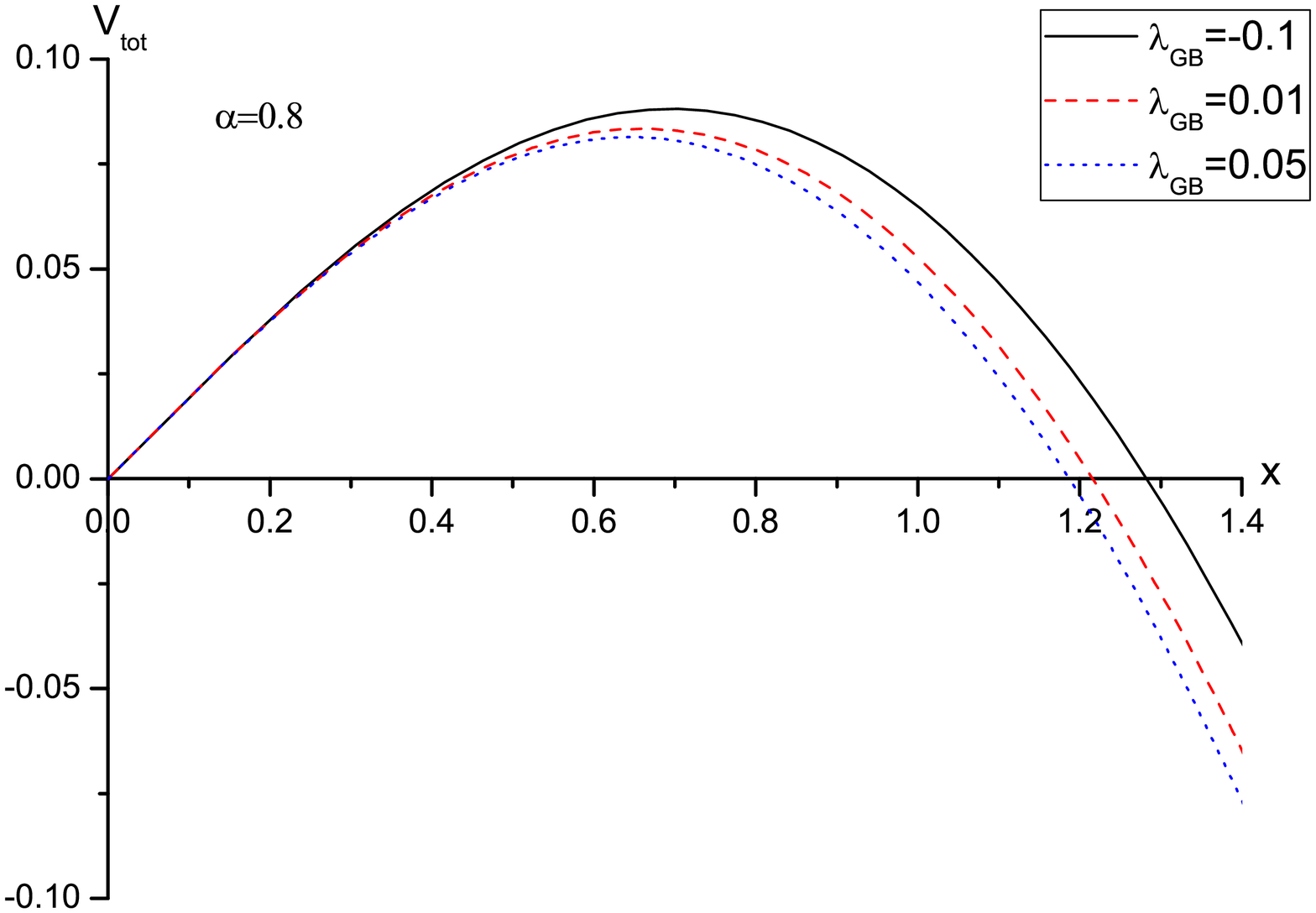}
\includegraphics[width=8cm]{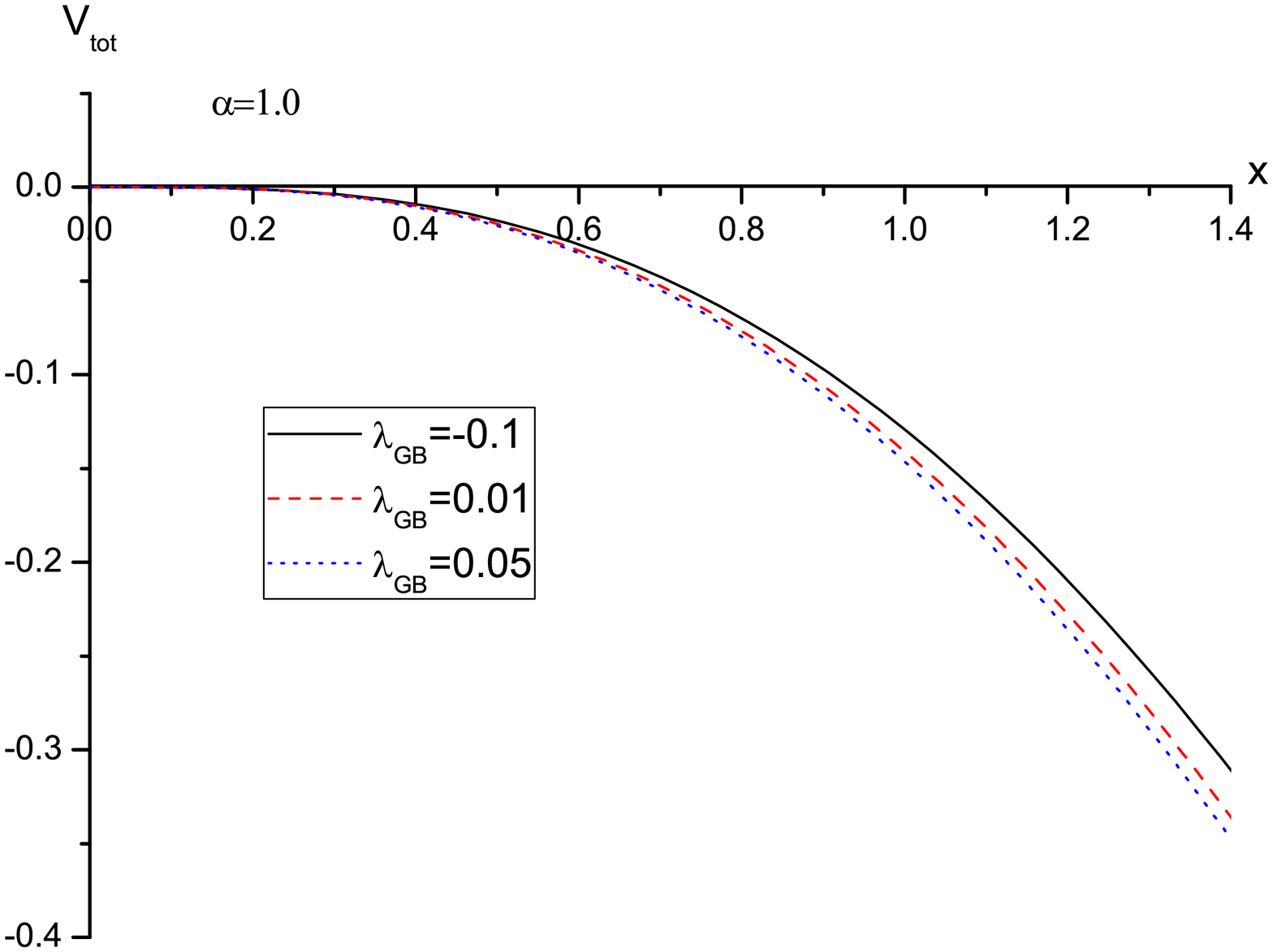}
\caption{$V_{tot}$ against $x$. Left: $\alpha=0.8$; Right:
$\alpha=1.0$. In all of the plots from top to bottom
$\lambda_{GB}=-0.1,0.01,0.05$, respectively.}
\end{figure}

To compare with the Einstein case in \cite{YS2}, we set $b=0.5$
and $T_Fr_0=L^2/r_0=1$. In Fig.1, we plot the total potential
$V_{tot}$ as a function of the inter-distance $x$ with two
different values of $\lambda_{GB}$, the left panel is for
$\lambda_{GB}=-0.1$ and the right one is for $\lambda_{GB}=0.05$.
In all of the plots from top to bottom $\alpha=0.8,0.9,1.0,1.1$,
respectively. From the figures, we can see that there is a
critical electric filed at $\alpha=1$ ($E=E_c$), in agreement with
\cite{YS2}.

To see the effect of $R^2$ corrections on the potential barrier,
we plot $V_{tot}$ against $x$ at $\alpha=0.8$ with different
values of $\lambda_{GB}$ in the left panel of Fig.2. We can see
that increasing $\lambda_{GB}$ leads to decreasing the height and
the width of the barrier. As we know, the higher the barrier, the
harder the produced pair escapes to infinity. Therefore, we
conclude that by increasing $\lambda_{GB}$ the Schwinger effect is
enhanced.

Moreover, to show the effect of $R^2$ corrections on $E_c$, we
plot $V_{tot}$ against $x$ at $\alpha=1$ with different
$\lambda_{GB}$ in the right panel of Fig.2. It is found that the
barrier vanishes for each plot implying the vacuum becomes
unstable. In fact, that the barrier of each plot disappears at
$\alpha=1$ can be strictly proved, i.e, we can calculate the
derivative of $V_{tot}$ at $x=0$ as,
\begin{equation}
\frac{dV_{tot}}{dx}\big{|}_{x=0}=(1-\alpha)mT_F\sqrt{f(r_0)}.
\end{equation}

\section{confining D3-brane background}

In this section we analyze the Schwinger effect in a confining
D3-brane background with $R^2$ corrections. The metric is given by
\cite{RG}
\begin{equation}
ds^2=-\frac{r^2}{L^2}dt^2+\frac{r^2}{L^2}(dx^1)^2+\frac{r^2}{L^2}(dx^2)^2+m^2f(r){\frac{r^2}{L^2}}(dx^3)^2+\frac{L^2}{r^2}\frac{dr^2}{f(r)}
\label{metric1},
\end{equation}
with
\begin{equation}
f(r)=\frac{1}{2\lambda_{GB}}\big{[}1-\sqrt{1-4\lambda_{GB}(1-\frac{r_h^4}{r^4})}\big{]}\label{f1},
\end{equation}
and
\begin{equation}
m^2=\frac{1}{2}(1+\sqrt{1-4\lambda_{GB}}),
\end{equation}
where $r_h$ is the inverse compactification radius in the $x^3$-
direction.

Similar to the previous section, we call again the inter-distance
and the sum of potential energy and static energy as $x$ and
$V_{PE+SE}$, respectively. One finds
\begin{equation}
x=\frac{2L^2}{r_0a}\int_1^{1/a}\frac{dy}{y^2\sqrt{(y^4-1)f_1(y)}}\label{xx1},
\end{equation}
and
\begin{equation}
V_{PE+SE}=2T_Fr_0a\int_1^{1/a}\frac{y^2dy}{\sqrt{(y^4-1)f_1(y)}},\label{en1}
\end{equation}
where $f_1(y)$ is defined in (\ref{f1y}).

The next step is to evaluate the critical electric field. Using
(\ref{metric1}), we have the induced metric as
\begin{equation}
G_{00}=-\frac{r^2}{L^2}, \qquad G_{11}=\frac{r^2}{L^2},\qquad
G_{22}=\frac{r^2}{L^2},\qquad G_{33}=m^2f(r)\frac{r^2}{L^2}.
\end{equation}

Then we get
\begin{equation}
G_{\mu\nu}+\mathcal{F}_{\mu\nu}=\left(
\begin{array}{cccc}
 -\frac{r^2}{L^2} & 2\pi\alpha^\prime E & 0 & 0\\
 -2\pi\alpha^\prime E & \frac{r^2}{L^2} & 0 & 0 \\
 0 & 0 & \frac{r^2}{L^2} & 0\\
 0 & 0 & 0 & m^2f(r)\frac{r^2}{L^2}
\end{array}
\right),
\end{equation}
which yields
\begin{equation}
det(G_{\mu\nu}+\mathcal{F}_{\mu\nu})=-m^2f(r)\frac{r^4}{L^4}[\frac{r^4}{L^4}-(2\pi
\alpha')^2E^2],\label{det11}
\end{equation}
where the electric field $E$ is turned on along the
$x^1$-direction as well.

Substituting (\ref{det11}) into (\ref{dbi1}) and setting the
D3-brane at $r=r_0$, one has
\begin{equation}
S_{DBI}=-mT_{D3}\frac{r_0^4}{L^4}\int d^4x
\sqrt{f(r_0)[1-\frac{(2\pi\alpha^\prime)^2L^4}{r_0^4}E^2]}\label{dbi10}.
\end{equation}
Obviously,
\begin{equation}
f(r_0)>0.
\end{equation}

To avoid (\ref{dbi10}) being ill-defined, one needs only
\begin{equation}
1-\frac{(2\pi\alpha^\prime)^2L^4}{r_0^4}E^2\geq0,
\end{equation}
results in
\begin{equation}
E\leq T_F\frac{r_0^2}{L^2}.
\end{equation}

Therefore, the critical field $E_c$ in the confining D3-brane
background with $R^2$ corrections is obtained
\begin{equation}
E_c=T_F\frac{r_0^2}{L^2}.
\end{equation}
one can see that in this case $E_c$ is not affected by the $R^2$
corrections. This is because the electric field is turned on
$x^1$-direction while the $x^3$-direction, related to
$\lambda_{GB}$, is compactified \cite{YS1}.

Likewise, the total potential is
\begin{eqnarray}
V_{tot}&=&V_{PE+SE}-Ex\nonumber\\&=&2T_Fr_0a\int_1^{1/a}\frac{y^2dy}{\sqrt{(y^4-1)f_1(y)}}-\frac{2T_F\alpha
r_0}{a}\int_1^{1/a}\frac{dy}{y^2\sqrt{(y^4-1)f_1(y)}}\label{V1},
\end{eqnarray}
where $\alpha$ is defined in (\ref{afa1}).

\begin{figure}
\centering
\includegraphics[width=8cm]{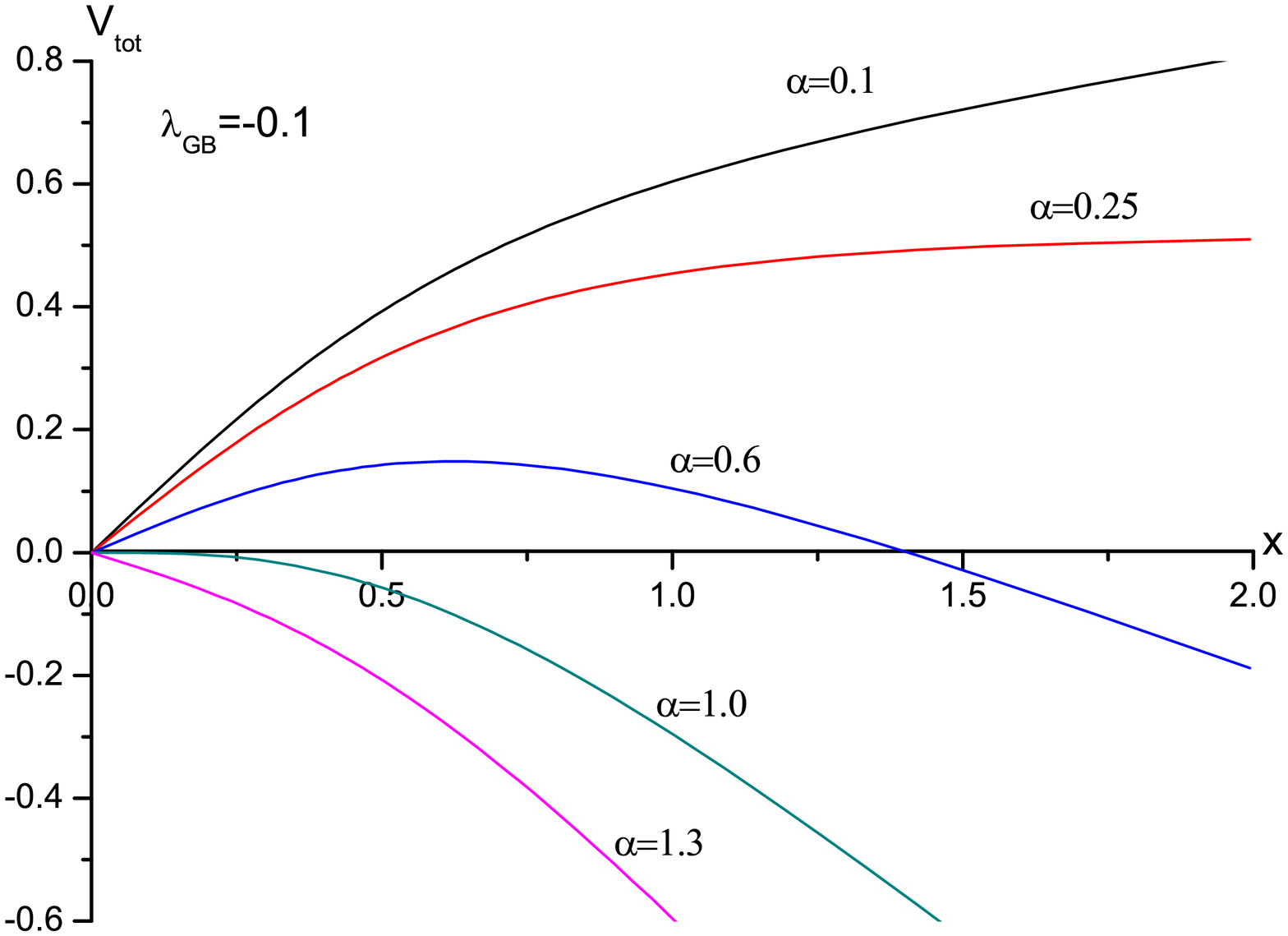}
\includegraphics[width=8cm]{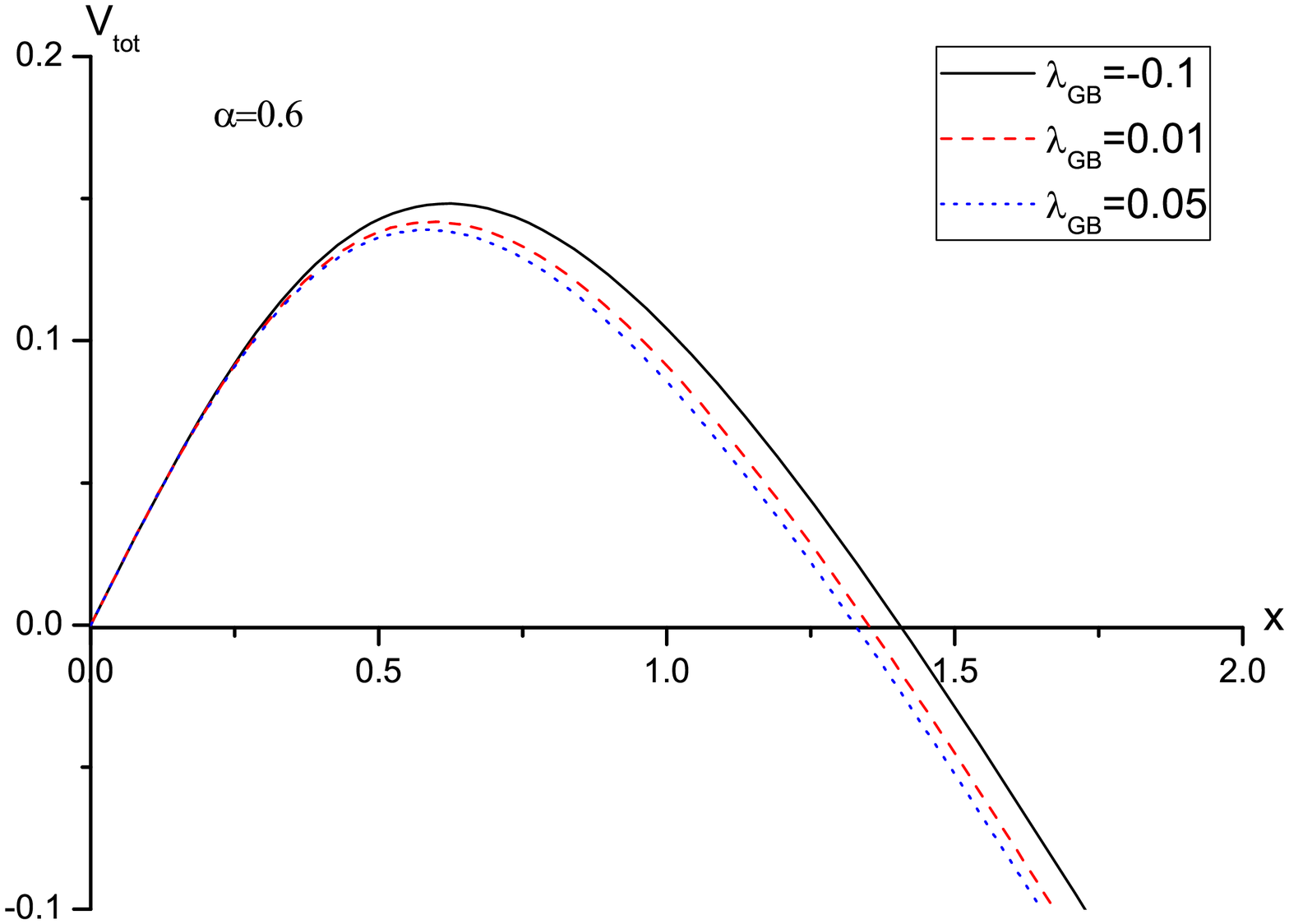}
\caption{$V_{tot}$ against x.  Left: $\lambda_{GB}=-0.1$, from top
to bottom $\alpha=0.1,0.25,0.6,1.0,1.3$; Right: $\alpha=0.6$, from
top to bottom $\lambda_{GB}=-0.1,0.01,0.05$, respectively. }
\end{figure}

\begin{figure}
\centering
\includegraphics[width=8cm]{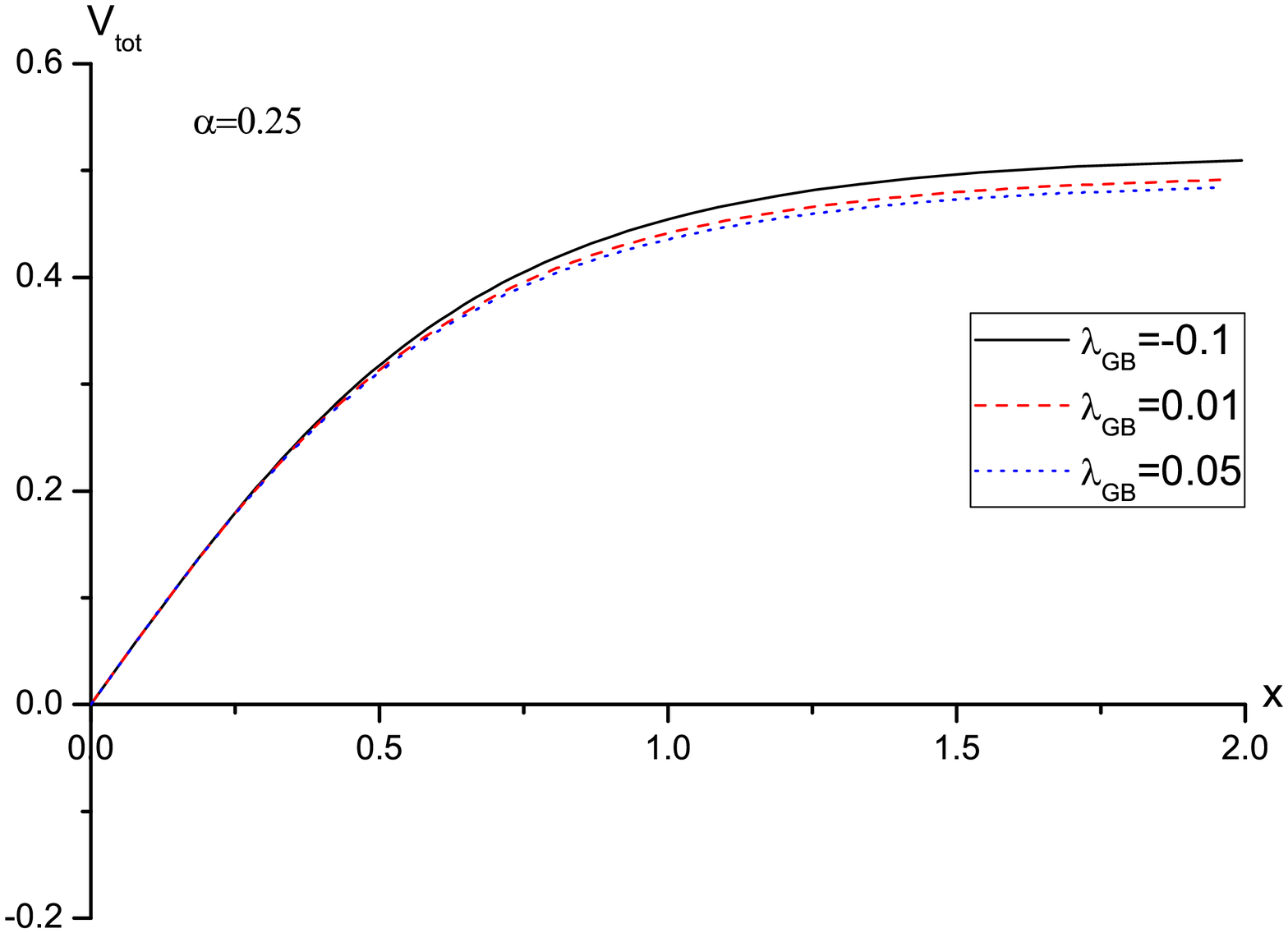}
\includegraphics[width=8cm]{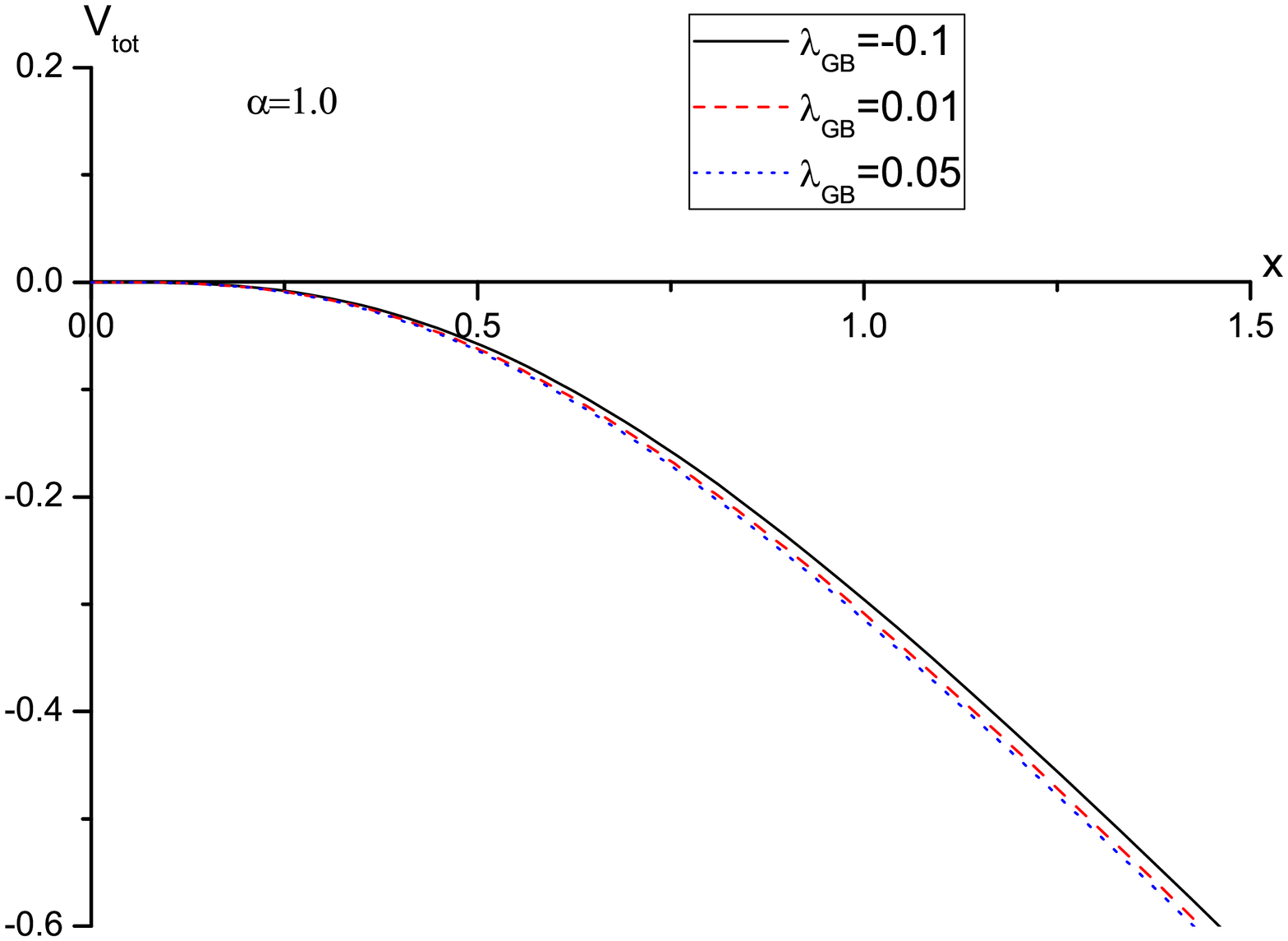}
\caption{$V_{tot}$ against $x$. Left: $\alpha=0.25$; Right:
$\alpha=1$. In all of the plots from top to bottom
$\lambda_{GB}=-0.1,0.01,0.05$, respectively. }
\end{figure}

According to the analysis of \cite{YS1}, there are two critical
values for the electric field in the confining D3-brane
background, $E_c=\frac{T_Fr_0^2}{L^2}$ and
$E_s=\frac{T_Fr_h^2}{L^2}$. When $E>E_c$, the potential barrier
vanishes and no tunneling occurs implying the vacuum becomes
unstable. When $E_s<E<E_c$, the potential barrier is present and
the Schwinger effect can be described as a tunneling process. When
$E<E_s$, the potential tends to diverge at infinitely and no
Schwinger effect occurs.

Let us discuss results. In the left panel of Fig.3, we plot
$V_{tot}$ versus $x$ with $\lambda_{GB}=-0.1$ by setting $b=0.5$
and $2L^2/{r_0}=2T_Fr_0=1$, as follows from \cite{YS1}. Other
cases with different $\lambda_{GB}$ have similar picture. From the
figures, one can see that there indeed exist two critical values
for the electric field: one is at $\alpha=1$($E=E_c$), the other
is at $\alpha=0.25$ ($E=E_s=0.25E_c$).

Likewise, to study $R^2$ corrections to the potential barrier, we
plot $V_{tot}$ versus $x$ at $\alpha=0.6$ with different
$\lambda_{GB}$ in the right panel of Fig.3. One can see that as
$\lambda_{GB}$ increases both the height and width of the
potential barrier decrease. Thus, one concludes that increasing
$\lambda_{GB}$ enhances the Schwinger effect, consistently with
the findings of \cite{SZ}.

Also, to show the effect of $R^2$ corrections on the two critical
electric fields, we plot $V_{tot}$ versus $x$ at $\alpha=0.25$ and
$\alpha=1$ with different $\lambda_{GB}$ in Fig.4. From the left
panel, one can see that by varying $\lambda_{GB}$ the potential
always becomes flat at $\alpha=0.25$, which means the critical
field $E_s$ is not modified by $R^2$ corrections. This is
consistent with the value of $E_s$ defined as
$E_s=\frac{T_Fr_h^2}{L^2}$. From the right panel, one finds that
the barrier vanishes for each plot at $\alpha=1$, in agreement
with the DBI result.

\section{conclusion and discussion}

In this paper, we have investigated $R^2$ corrections to the
holographic Schwinger effect in an AdS black hole background and a
confining D3-brane background, respectively. The critical values
for the electric field were obtained. It is shown that for both
backgrounds increasing the Gauss-Bonnet parameter the Schwinger
effect is enhanced. Moreover, the critical electric field $E_c$ is
dependent on $\lambda_{GB}$ in the AdS black hole background but
not affected by it in the confining D3-brane background.

In addition, the results may provide an estimate of how the
Schwinger effect changes with $\eta/s$ at strong coupling. From
(\ref{eta}) one knows that increasing $\lambda_{GB}$ leads to
decreasing the $\eta/s$ thus making the fluid becomes more
"perfect". On the other hand, increasing $\lambda_{GB}$ leads to
increasing the Schwinger effect. Therefore, one concludes that at
strong coupling as $\eta/s$ decreases the Schwinger effect is
enhanced.

Finally, we should admit that we cannot predict a result for
$\mathcal N=4$ SYM theory because the first higher derivative
correction is related to $R^4$ terms but not $R^2$. We leave this
for further study.

\section{Acknowledgments}

The authors would like to thank the anonymous referee for his/her
valuable comments and helpful advice. This work is partly
supported by the Ministry of Science and Technology of China
(MSTC) under the ¡°973¡± Project no. 2015CB856904(4). Zi-qiang
Zhang and Gang Chen are supported by the NSFC under Grant no.
11475149. De-fu Hou is supported by the NSFC under Grant no.
11375070 and 11521064.



\begin{thebibliography}{0}

\bibitem{JS}
J.S. Schwinger, "{\sl On gauge invariance and vacuum
polarization,"} Phys. Rev. 82 (1951) 664.

\bibitem{IK}
I.K. Affleck and N.S. Manton, {\sl " Monopole pair production in a
magnetic field,"} Nucl. Phys. B 194, 38 (1982).

\bibitem{Maldacena:1997re}
J.M. Maldacena, {\sl "The large $N$ limit of superconformal field
theories and supergravity,"} Adv. Theor. Math. Phys. 2, 231
(1998).

\bibitem{Gubser:1998bc}
S.S. Gubser, I.R. Klebanov and A.M. Polyakov, {\sl "Gauge theory
correlators from non-critical string theory,"}  Phys. Lett. B428,
105 (1998) [hep-th/9802109].

\bibitem{MadalcenaReview}
O. Aharony, S.S. Gubser, J. Maldacena, H. Ooguri and Y. Oz, {\sl
"Large N field theories, string theory and gravity,"} Phys. Rept.
{\bf 323}, 183 (2000).


\bibitem{GW}
G.W. Semenoff and K. Zarembo, {\sl "Holographic Schwinger
effect,"} Phys. Rev. Lett. 107 (2011) 171601 [hep-th/1109.2920].


\bibitem{YS}
Y. Sato and K. Yoshida, {\sl " Universal aspects of holographic
Schwinger effect in general backgrounds,"} JHEP 1312 (2013) 051
[hep-th/1309.4629].

\bibitem{YS1}
Y. Sato and K. Yoshida, {\sl "Holographic Schwinger effect in
confining phase,"} JHEP 1309 (2013) 134 [hep-th/1306.5512].

\bibitem{YS2}
Y. Sato and K. Yoshida, {\sl "Potential Analysis in Holographic
Schwinger Effect,"} JHEP 1308 (2013) 002.


\bibitem{SB}
S. Bolognesi, F. Kiefer and E. Rabinovici, {\sl "Comments on
critical electric and magnetic fields from holography,"} JHEP 1301
(2013) 174 [hep-th/1210.4170].


\bibitem{KH}
K. Hashimoto, T. Oka and A. Sonoda,{\sl "Electromagnetic
instability in holographic QCD,"} [hep-th/1412.4254].

\bibitem{JS1}
J. Sadeghi, B. Pourhassan, S. Tahery and F. Razavi,{\sl
"Electrostatic potential in the holographic Schwinger effect with
a deformed AdS background,"} [hep-th/1603.07629].


\bibitem{SZ}
S.J. Zhang and E. Abdalla,{\sl "Gauss-Bonnet corrections to
holographic Schwinger effect in confining
background,"}Gen.Rel.Grav. 48 (2016) no.5, 60 [hep-th/1508.03364].


\bibitem{JA}
J. Ambjorn, Y. Makeenko,{\sl "Remarks on Holographic Wilson Loops
and the Schwinger Effect,"} Phys.Rev.D 85, 061901 (2012)
[hep-th/1112.5606].

\bibitem{KH1}
K. Hashimoto, T. Oka,{\sl "Vacuum Instability in Electric Fields
via AdS/CFT: Euler-Heisenberg Lagrangian and Planckian
Thermalization,"} JHEP 10 (2013) 116 [hep-th/1307.7423].

\bibitem{DD}
D.D. Dietrich,{\sl "Worldline holographic Schwinger effect,"}
Phys. Rev. D 90, 045024 (2014) [hep-ph/1405.0487].

\bibitem{WF}
W. Fischler, P. H. Nguyen, J. F. Pedraza, W. Tangarife,{\sl
"Holographic Schwinger effect in de Sitter space,"} Phys. Rev. D
91, 086015 (2015) [hep-th/1411.1787].

\bibitem{MG}
M. Ghodrati,{\sl "Schwinger Effect and Entanglement Entropy in
Confining Geometries,"} Phys. Rev. D 92, 065015 (2015)
[hep-th/1506.08557].

\bibitem{XW}
X. Wu,{\sl "Notes on holographic Schwinger effect,"} JHEP 09
(2015) 044 [hep-th/1507.03208].

\bibitem{SC}
S. Chakrabortty and B. Sathiapalan,{\sl "Schwinger effect and
negative differential con-ductivity in holographic models,"} Nucl.
Phys. B 890 (2014) 241 [hep-th/1409.1383].

\bibitem{ZQ1}
Z.q. Zhang, D.f. Hou, Y. Wu and G. Chen,  {\sl "Holographic
Schwinger effect in a confining D3-brane background with chemical
potential"}, Advances in High Energy Physics, Volume 2016 (2016),
Article ID 9258106.

\bibitem{AS}
A.S. Gorsky, K.A. Saraikin and K.G. Selivanov, {\sl "Schwinger
type processes via branes and their gravity duals,"} Nucl. Phys. B
628 (2002) 270 [hep-th/0110178].

\bibitem{DK1}
D. Kawai, Y. Sato and K. Yoshida, {\sl "A holographic description
of the Schwinger effect in a confining gauge theory,"} Int. J.
Mod. Phys. A 30, 1530026 (2015) [hep-th/1504.00459].

\bibitem{MRD}
M.R. Douglas and S. Kachru, {\sl "Flux compactification,"} Rev.
Mod. Phys. 79 (2007) 733 [hep-th/0610102].


\bibitem{MB}
M. Brigante, H. Liu, R.C. Myers, S. Shenker, S.Yaida {\sl
"Viscosity Bound Violation in Higher Derivative Gravity"},
Phys.Rev.D {\bf 77} 126006(2008). [hep-th/0712.0805].

\bibitem{MB1}
M. Brigante, H. Liu, R.C. Myers, S. Shenker, and S. Yaida, {\sl
"The Viscosity Bound and Causality Violation"}, Phys. Rev. Lett.
100, 191601 (2008).

\bibitem{YK}
Y. Kats and P. Petrov, {\sl "Effect of curvature squared
corrections in AdS on the viscosity of the dual gauge theory"},
JHEP. 01 (2009) 044.

\bibitem{JN}
J. Noronha and Adrian Dumitru, {\sl"Heavy quark potential as a
function of shear viscosity at strong coupling"}, Phys. Rev. D
{\bf 80} 014007 (2009) [hep-ph/0903.2804].

\bibitem{JN1}
S.I. Finazzo and J. Noronha, {\sl"Estimates for the Thermal Width
of Heavy Quarkonia in Strongly Coupled Plasmas from Holography"},
JHEP {\bf 11} 042 (2013). [hep-th/1306.2613].


\bibitem{KB1}
K.B. Fadafan,  {\sl "$R^2$ curvature-squared corrections on drag
force"}, JHEP {\bf 0812} 051 (2008). [hep-th/0803.2777].


\bibitem{ZQ2}
Z.q. Zhang, D.f. Hou, Y. Wu and G. Chen,  {\sl "$R^2$ corrections
to the jet quenching parameter"}, Advances in High Energy Physics,
Volume 2016 (2016), Article ID9503491.

\bibitem{PK}
P.Kovtun, D.T. Son, A.O. Starinets, {\sl"Viscosity in Strongly
Interacting Quantum Field Theories from Black Hole Physics"},
Phys. Rev. Lett {\bf 94} 111601 (2005) [hep-th/0405231].

\bibitem{SM}
S.M. Carroll, {\sl "Spacetime and geometry: An introduction to
general relativity"}, San Francisco, USA: Addison-Wesley (2004)
513 p

\bibitem{BZ1}
B. Zwiebach, {\sl "Curvature squared terms and string theories"},
Phys. Lett. 156B, 315 (1985).

\bibitem{DM}
D.M. Hofman, J. Maldacena, {\sl "Conformal collider physics:
Energy and charge correlations"}, JHEP 0805 012 (2008)
[hep-th/0803.1467].


\bibitem{RG}
R.G. Cai, {\sl "Gauss-Bonnet black holes in AdS spaces"}, Phys.
Rev. D 65, 084014 (2002) [hep-th/0109133].









\end{thebibliography}
\end{document}